# Relativity, GPS, and the Validity of Common View Synchronization

Revised 3/16/2012 09:18:00

Eric L. Michelsen[a)]

*University of California, San Diego, Department of Physics*
*9500 Gilman Dr. MS-0424 SERF-317, La Jolla, CA, USA 92093*

We show that Common View Synchronization is valid to synchronize distant clocks. We describe the relativistic physics, noting that a prerequisite for synchronization is the existence of a space-time with a stationary metric for each clock. The analysis shows that there are no Special Relativistic effects that need be included in the synchronization method (beyond those normally used in a single GPS-based earth clock). In particular, synchronizing a ground clock to the GPS satellite does *not* make that clock keep time in the reference frame of the satellite. Symmetries are very helpful in analyzing the behavior. We briefly describe some practical considerations in synchronizing distant earth clocks, such as antenna cabling and variations in receiver electronics, and how Common View Synchronization accommodates them.

## I. INTRODUCTION

Recently, the OPERA team's reporting of faster-than-light neutrinos[1] has brought attention to the clock synchronization method used to measure the time of flight (TOF). The launch time is measured by a clock at CERN, and the detection time is measured by a different clock at Gran Sasso, 730 km away. Since different clocks are used for launch and detection, they must be carefully synchronized.

The OPERA team uses Common View Synchronization (CVS) to synchronize the two clocks. In essence, CVS works by having both ground clocks synchronize to the *same* Global Positioning System (GPS) satellite at the same time.[2] This cancels most of the atmospheric delays, since the atmosphere is similar, even over a distance of 730 km. However, the procedure is complicated by the need to calibrate out differences in cabling and electronics between the antennas and the digital outputs (pulse-per-second or PPS). This calibration requires a portable GPS clock be carried from one clock to the other.

Some recent postings[3,4] proposed that CVS induces a time difference between the two ground clocks, which would cause inaccurate TOF measurements. We show here that CVS, in fact, works well, and has no inherent error. It also does not require any compensation for Special Relativity (SR) (beyond those normally used in a single GPS-based earth clock).

We first discuss the relativistic considerations in general clock synchronization. We then describe the CVS procedure in use. We address here only the fundamental physics involved, and ideal circular orbits. There are many other practical considerations which must be resolved in a real system.[6]

## II. RELATIVISTIC ANALYSIS

Before discussing any form of synchronization in which relativistic effects are important, it is critical to define both the reference frames involved, and the coordinate systems used. Many an analysis has gone wrong because of subtle shifts in these definitions.

Synchronizing multiple clocks requires two separate steps: rate synchronization, and epoch synchronization. Rate synchronization achieves the clocks counting time at the same rate, as defined by a particular observer, aka reference frame. Epoch synchronization achieves a common time origin, as measured by the same observer, i.e. in the same reference frame as that used for rate synchronization. In the case of GPS disciplined clocks, rate synchronization requires consideration of both SR and GR effects. However, symmetries in the system greatly simplify these considerations. We now discuss each synchronization step in detail.

For there to be any reasonable chance of rate synchronization, there must exist a reference frame in which each clock has a stationary metric *throughout its motion*. Our first job, then, is to establish the existence of such a frame, which must include the two earth clocks and the GPS satellite clock.

Before considering the real system, consider a general system comprising clocks orbiting a central point (Fig. 1). The nature of the orbit is not important here; it may be a free-falling satellite, or a clock fixed on a rotating rigid body. The central clock may be inside a massive body, with significant gravitational potential at the orbiting clock radius.

The center of the system is the point of maximal symmetry, and is therefore a good candidate for a reference frame and coordinate system.[5] Assume a non-rotating, inertial central observer. (Non-rotating with respect to what? The distant stars! A full discussion of this choice is outside our scope, but the equatorial bulge of the earth provides empirical evidence for its validity. Note that even here we must define our "non-rotating" reference frame.) Consider a ring of clocks, all orbiting at the same radius and constant angular velocity. These clocks transmit timing signals to the central observer. By symmetry, all orbiting clocks run at the same rate, as measured by the central observer. They may run at a different rate than the central clock, but they can be made to be synchronized to the central clock by speeding or slowing their local time rate so that the central observer measures them at the same rate as her local clock.

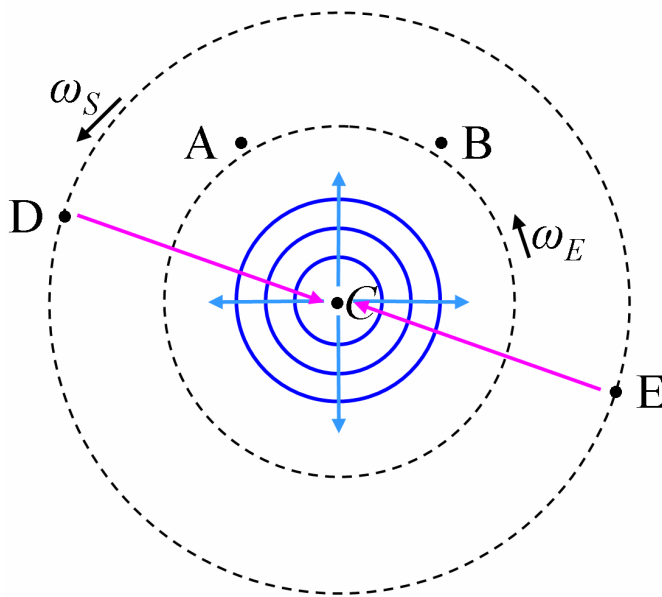

FIG. 1. (Color online) A central clock, orbited by one or more clocks. The arrows indicate time signal transmissions.

Such rate-synchronization can be done for any orbit, any angular velocity, and any orbital plane. Furthermore, the central observer can echo timing signals from any orbiting clock to any other, and therefore all clocks, in all circular orbits, can be simultaneously rate-synchronized, as measured by the central observer. In this case of orbiting clocks and a central clock, rate synchronization is transitive: if A synchronizes to C, and B synchronized to C, then A synchronizes to B. (There are other systems where rate synchronization is not transitive, but they do not concern us here.)

A crucial point is that in this reference frame, there is no need to consider any SR effect *between* orbiting clocks, even though they are in changing relative motion.

Next, we can achieve epoch synchronization in our central reference frame, by transmitting labeled timing pulses from the center to the orbiting clocks. Since the metric is stationary, the distance from the center to any clock is constant in time. The transmitted signals travel at a known speed, and the orbiting clocks know their radii, so they can adjust for the propagation delay from the center, as measured in the inertial central frame. In this way, all clocks, at all radii and angular velocities, can be simultaneously epoch-synchronized, in the global inertial frame. Furthermore, the system of synchronized clocks can also be used to define a global time coordinate for a *rotating* reference frame at arbitrary angular velocity. Such a reference frame is *not* inertial, and is therefore subject to the Sagnac effect.

The actual system in the neutrino TOF experiment consists of clocks on the earth, and orbiting satellite clocks. As just described, there exists a stationary reference frame in which all earth and satellite clocks can be simultaneously rate- and epoch- synchronized. This is exactly what the GPS system does. In particular, synchronizing an earth clock to a GPS clock does *not* make the earth clock run in the inertial frame of the satellite; it runs in the global inertial frame of the center of the earth. However, in this frame, the GPS system chooses proper time at the earth's *surface* as the time coordinate. Therefore, there is no need for any clock or observer at the earth's center. The system of clocks now also defines a global time coordinate for the *rotating* earth-surface reference frame, in which most experiments are actually performed.

Note that a standard clock at GPS altitude runs faster than clocks on the earth by 45 μs/day, because of the higher Newtonian gravitational potential. In addition, in the inertial earth-center frame, a GPS satellite clock is moving, and therefore slowed from SR time dilation by 7 μs/day. The net effect is that satellite clocks run faster than earth clocks by 38 μs/day. Actual GPS clocks are preset on the ground to run slowly by this amount, so that in orbit they closely match earth clock rates. (GPS clocks are also regularly updated from the ground, once or more per day.)

The neutrino TOF experiment is performed in the rotating earth-surface frame, which is not inertial. Therefore, the Sagnac effect contributes to measurements. However, in the neutrino experiment, the OPERA team has taken that into account, though the effect is negligible (2.2 ns).[1]

Why not use the satellite frame of reference? It is inertial, but only in an infinitesimal neighborhood around it. Still, the earth clocks are all at the same gravitational potential. However, in the satellite frame, there is no simplifying symmetry. The earth orbits the satellite, and the earth clocks have complicated motions and varying speeds, Clocks on one side of the earth have earth rotation motions that add to the orbital speed, while clocks on the opposite side of the earth have rotational motion that subtracts from the orbital speed. Varying speed imply that the clock rates (in the satellite frame) are not constant. In this frame, earth clocks are neither rate nor epoch synchronized, nor can they be. In principle, the analysis could be done this way, but it would be prohibitively complicated by these motions.

## III. COMMON VIEW SYNCHRONIZATION

We now apply the above results to the practical method for synchronizing earth clocks. The U.S. National Institute for Standards and Technology (NIST) defines a procedure for epoch synchronizing distant earth clocks from a single GPS satellite, called Common View Synchronization (CVS).[2] It works by having both earth clocks simultaneously receive a single GPS satellite (the satellite is in their "common view"). Simultaneous ranging to a single satellite causes atmospheric effects to largely cancel. CVS is widely used all over the world for many scientific measurements.

After epoch synchronization, each clock then runs at its own rate, but they can be held in close time synchronization by periodically rate synchronizing to the GPS system (not Common View). Rate adjustments to the earth clocks are deliberately slow and gradual, and thus the earth clock rate is a long-term average of the GPS rates, which is desirable.

If the two earth clocks were truly identical, then a single common view adjustment between them would fully epoch synchronize them. In practice, though, the two earth clocks cannot reasonably be made identical. Unavoidable differences include different cabling from the GPS antenna to the receiver electronics, and different parasitics and component variations in the electronics. These differences lead to a clock-specific time offset, relative to the global inertial reference frame. To eliminate these offsets, a full Common View Synchronization involves a portable GPS receiver/clock, and 3 steps:

1. Bring the portable clock antenna to earth clock A's antenna (say, within a few tens of cm). Use a common view of one satellite to synchronize earth clock A to the portable clock. Use the time-delta between clock A and the portable clock as an offset to clock A.
2. Transport the portable clock antenna to earth clock B's antenna, again within a few tens of cm. The portable clock need not keep time during transport, and may be powered off. Use a common view to synchronize clock B to the portable clock, again using the time-delta between clock B and the portable clock as an offset to clock B.
3. Use a common view to epoch synchronize clock A and clock B, using the time-deltas for each clock as previously determined from the portable clock.

Note that the portable clock may have its own time-delta, but since it is used to calibrate both earth clocks, any fixed offset in the portable clock is incorporated equally in the two earth clocks, thus achieving epoch synchronization of the two clocks with each other, in both the inertial earth frame, and the rotating earth-surface frame. That is all that is needed for TOF measurements.

Note that in each of steps 1 and 2, both clocks are "nearby," so any atmospheric delays are common, and do not appear in the time-delta between the two. However, many hours or days may elapse between steps 1 and 2, and the atmosphere may change appreciably in that time. That leads to a possible time offset between clocks A and B after step 2. Step 3 then establishes epoch synchronization, because even at 730 km separation, the atmospheric effects largely cancel (the satellite orbits at 20,200 km above the earth's surface).

## IV. CONCLUSIONS

The arguments here show that Common View Synchronization is valid to synchronize distant clocks, in both the inertial earth frame, and the rotating earth-surface frame. A prerequisite for any synchronization is the existence of a space-time with stationary metric, and clocks whose motion keeps their metrics stationary. There are no Special Relativistic effects that need be included in the synchronization method. In particular, synchronizing a ground clock to the GPS satellite does *not* make that clock keep time in the reference frame of the satellite; instead, it keeps time in the global frame of the earth center, and at a rate of standard clocks on the earth's surface. Symmetries are very helpful in analyzing the behavior of the clocks in the system. The 3-step CVS method accommodates the practical difficulties of varying antenna cabling and GPS receiver electronics between the two clocks, achieving a high-accuracy epoch synchronization of the clocks.

## 1. ACKNOWLEDGEMENTS

I thank Kim Griest for helpful discussions, and for reviewing the first written description of this analysis.

[a] Electronic mail: emichels@physics.ucsd.edu